\documentclass{article}

\usepackage[english]{babel}

\usepackage[letterpaper,top=2cm,bottom=2cm,left=3cm,right=3cm,marginparwidth=1.75cm]{geometry}

\usepackage{amsmath}
\usepackage{graphicx}
\usepackage[colorlinks=true, allcolors=blue]{hyperref}
\usepackage{comment}
\usepackage{authblk}

\title{Fraction Constraint in Partial Wave Analysis}
\author[1]{Xiang Dong}
\author[1]{Chu-Cheng Pan}
\author[1]{Yu-Chang Sun}
\author[1]{Ao-Yan Cheng}
\author[1]{Ao-Bo Wang}
\author[1]{Hao Cai\thanks{hcai@whu.edu.cn}}
\author[2]{Kai Zhu\thanks{zhuk@ihep.ac.cn}}
\affil[1]{Wuhan University, Wuhan 430072, People's Republic of China}
\affil[2]{Institute of High Energy Physics, Beijing 100049, China}

\begin{document}
\maketitle

\begin{abstract}
To resolve the non-convex optimization problem in partial wave analysis, this paper introduces a novel approach that incorporates fraction constraints into the likelihood function. This method offers significant improvements in both the efficiency of pole searching and the reliability of resonance selection within partial wave analysis.
\end{abstract}

\section{Introduction}
Partial wave analysis (PWA) is a powerful technique used in particle physics to study the angular distributions of particles produced in scattering or decay processes~\cite{klempt2017partial}. By decomposing the final-state wave functions into a sum of partial waves with different angular momentum quantum numbers, PWA allows people to extract valuable information about the underlying dynamics of the interaction\cite{peters2006primer,pilloni2017amplitude}. This method enables people to identify and study resonances, determine their properties such as masses and widths, and understand the contributing amplitudes and phase shifts. PWA is particularly useful in experiments involving complex final states or multiple particles, where it helps disentangle the different contributions and extract meaningful physical observables. PWA is widely employed in experiments involving hadron colliders, electron-positron colliders, and other facilities, making it an essential tool for studying the fundamental building blocks of matter and the forces that govern their interaction.

However, PWA usually suffers from non-convex optimization problems. Non-convexity arises due to the complex nature of the underlying physics and the presence of multiple resonances, therefore numerous undetermined parameters in a fitting model~\cite{berger2010partial}. Unlike convex optimization problems that have a unique global minimum, non-convex optimization problems have multiple local minima. This makes finding the best fit parameters challenging, as traditional optimization algorithms can get trapped in local minima and fail to find the global or near-global minimum. The non-convex nature of the problem introduces uncertainties and can lead to biased or inaccurate results. Overcoming these challenges requires the development and application of specialized non-convex optimization techniques that can effectively explore the parameter space and find the best fit solutions.

In this paper, we propose to mitigate the non-convex optimization problem in PWA by modifying the likelihood function with an additional penalty term. This term is related to a sum of all resonance state fractions. After introduce the definition of the additional penalty term, we perform two simplified PWAs, one is without the penalty term but the other one with, on a toy Monte Carlo (MC) sample. General features are obtained for the proposed PWA method, and compared with the conventional one. Then we discuss how to obtain a crucial parameter in the penalty term by a scanning method, that is more practical in a real measurement than the previously pedagogical one. Meanwhile, we show the proposed method is helpful to select reasonable contributions of resonances. A short summary then ends this paper.

\section{Fraction Constraints to the Partial Wave Analysis}
As mentioned in the introduction, there are usually many undetermined parameters in a PWA, so the fitting is essentially a non-convex optimization problem, that will result in a non-global minimum point, sometimes as an unreasonable result. To resolve this problem, we propose to add a penalty term to the traditional logarithm of the likelihood, $-\ln L$, to construct a new target function $\tilde{M}$:
\begin{equation}
\tilde{M}=-\ln L+\lambda(\mathbf{SF}-\overline{\mathbf{SF}})^2 \ ,
\end{equation}
where $\mathbf{SF}$ is the sum of the fractions of total events as $\mathbf{SF}=\sum_k\mathbf{F}_k$ and $\overline{\mathbf{SF}}$ is its expected value, where $k$ is the index of the amplitude, and $\lambda$ is the strict-factor. The determination of $\overline{\mathbf{SF}}$ and $\lambda$ are based on the situations that will be discussed later. Explicitly, the fraction of each channel is defined as:
\begin{equation}
\mathbf{F}_k=\frac{1}{N}\sum_{i=1}^N \frac{\left|c_k M_k\left(\zeta_i\right)\right|^2}{\left|\sum_k c_k e^{\mathrm{i} \phi_k} M_k\left(\zeta_i\right)\right|^2} \ ,
\end{equation}
where $N$ is the number of events, $M_k$ are the (normalized) amplitude with respect to $\zeta_i$ representing both physical and nuisance parameters that may dynamically depend on the $i$th event, $c_k$ and $\phi_k$ are the magnitude and phase of each amplitude. By introducing this additional term, we restrict the feasible region and transform the original optimization problem into a “constrained non-convex optimization”, that potentially is more tractable.  Here, $\overline{\mathbf{SF}}$ is the expected value of $\mathbf{SF}$. Since $\mathbf{SF}$ represents only the contribution from non-interference effect, the value of $\mathbf{SF}$ is usually not 100\%. When constructive interference dominates between resonance states, $\mathbf{SF}$ will be less than 100\%; when destructive interference dominates between resonance states, $\mathbf{SF}$ will be greater than 100\%. But no matter the interference is constructive or destructive, we expect the $\mathbf{SF}$ based on a reasonable physical solution will not extremely deviate from 100\%. Obviously, when $\lambda$ is close to zero, $\tilde{M}$ will be reduced to $-\ln L$; but when $\lambda$ is large enough, $\mathbf{SF}$ will be restricted to $\overline{\mathbf{SF}}$, i.e., the interference effect is under control, then the parameter space will be deduced, and the convexity is improved. 

\section{Partial Wave Analysis without or with Fraction Constraints}
\begin{figure}[htbp]
  \centering
  \includegraphics[width=0.6\textwidth]{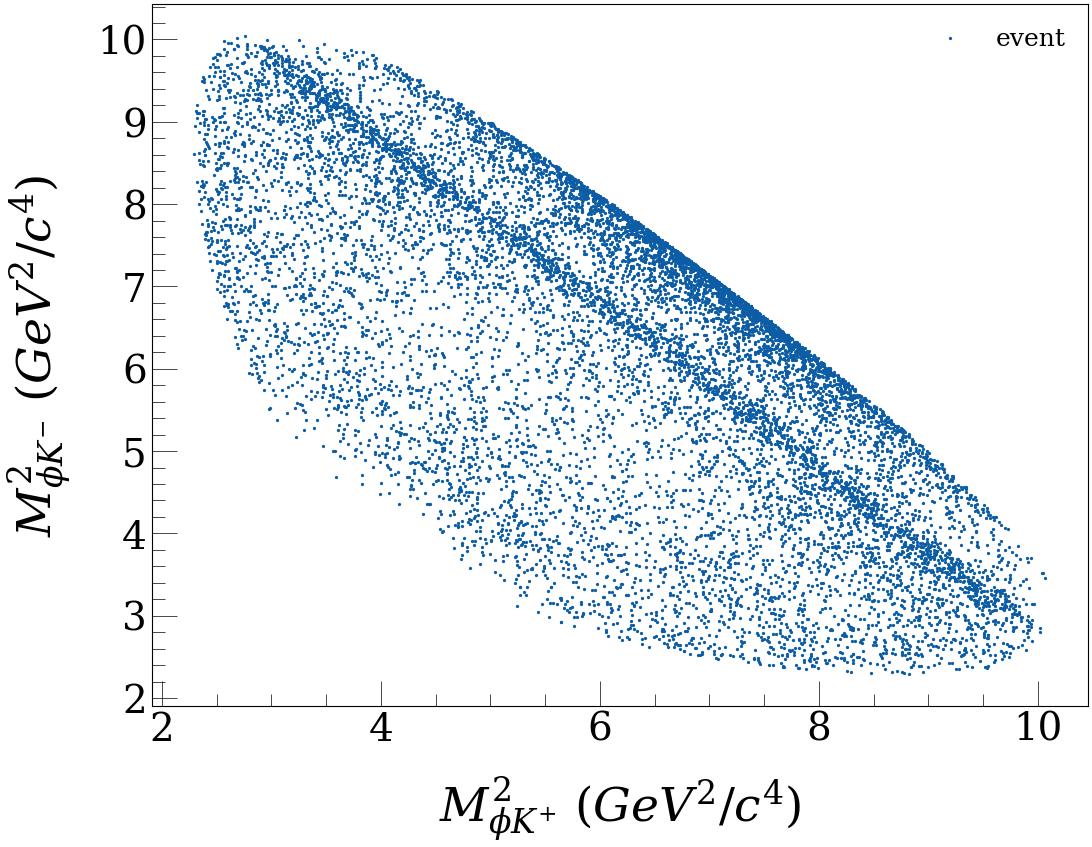} 
  \caption{The Dalitz plot from the MC sample $S_0$ generated by the $R_0$ model.}
  \label{fig-1a}
\end{figure}

\begin{figure}[htbp]
  \centering
  \includegraphics[width=0.6\textwidth]{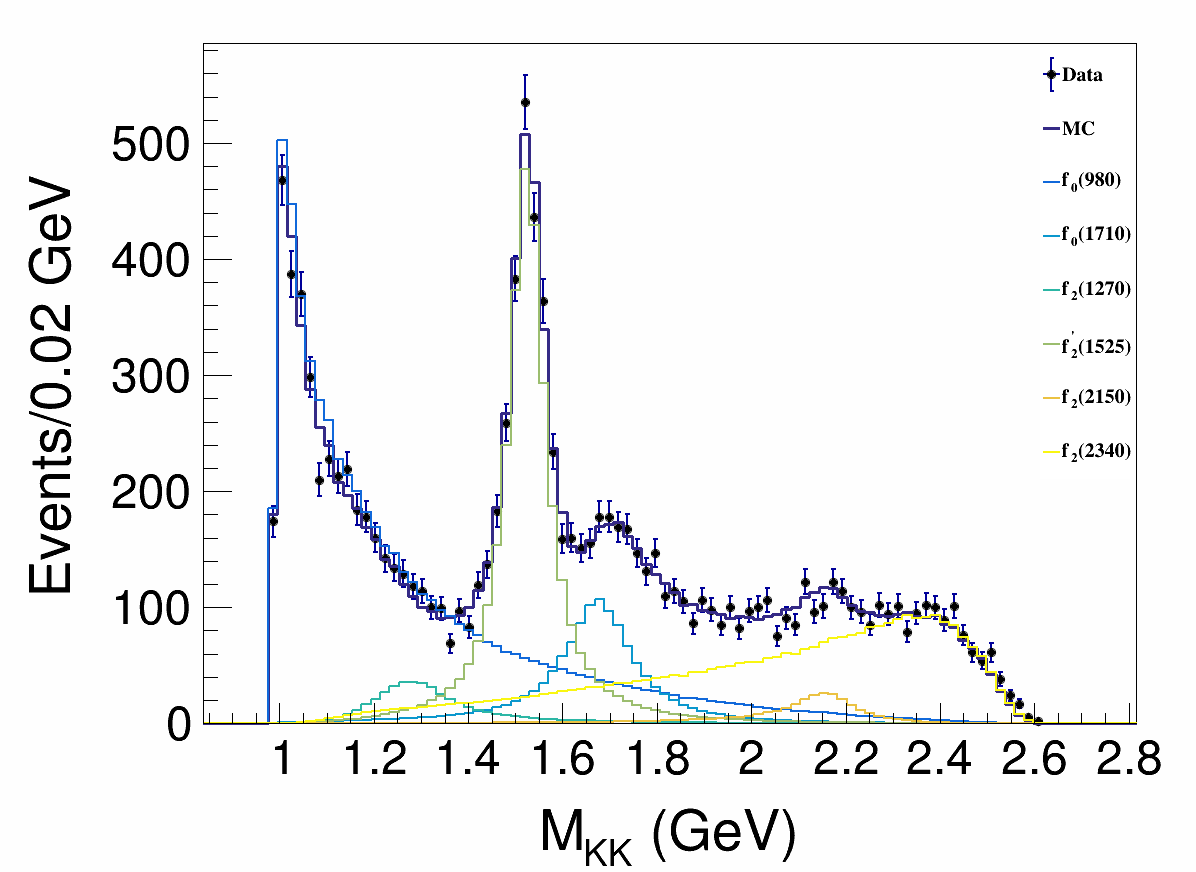} 
  \caption{The $K^+ K^-$ invariant mass spectrum for the MC sample $S_0$ generated by the $R_0$ model.}
  \label{fig-1b}
\end{figure}

For demonstration, an MC sample containing 10,000 events have been generated based on a PWA model that describes the process $\psi(2S) \rightarrow \phi K^{+} K^{-}$~\cite{ZouBS} with various intermediate resonances decaying into $K^+K^-$. For convenience, this PWA model is denoted as $R_0$ and the MC sample is denoted as $S_0$. In $R_0$, resonances such as $f_0(980)$~\cite{flatte1,flatte2}, $f_2(1270)$~\cite{Shchegelsky:2006et}, $f_2^{\prime}(1525)$~\cite{Longacre:1986fh}, $f_0(1710)$~\cite{Belle:2013eck}, $f_2(2150)$~\cite{WA102:2000lao}, and $f_2(2340)$~\cite{BESIII:2022zel} are included with description according to the corresponding references, respectively. Their masses, widths, and relevant fractions are shown in Table~\ref{tab-1}. In the $R_0$ model, covariant tensors are applied to describe the partial wave amplitudes. It should be noted that Table~\ref{tab-1} lists the fractions of each resonance, and the sum of the fractions yields a $\mathbf{SF}$ value of approximately 115\%. The Dalitz plot corresponding to the generated events is shown in Fig.~\ref{fig-1a}, and the distribution on the $K^{+} K^{-}$ invariant mass spectrum is shown in Fig.~\ref{fig-1b}. The existence of both  narrow and broad resonances makes $R_0$ not a na\"{i}ve model. It should be noted that this MC sample is just designed for studying the PWA method, but does not intend to simulate the three-body decay $\psi(2S) \rightarrow \phi K^{+} K^{-}$ in the real world.

\begin{table}[htbp]
\caption{Resonances incorporated in PWA model $R_0$, and their corresponding parameters.}
\centering
\begin{tabular}{|c|c|c|c|c|}
\hline
$R_0$ & Name & $F_i$(\%) & Mass (GeV) & Width (GeV)\\ \hline
1 & $f_{0}(980)$ & 39.5 & 0.979 & 0.107 \\ 
2 & $f_{2}(2340)$ & 37.1 & 2.548 & 0.324 \\ 
3 & $f^{'}_{2}(1525)$ & 24.7 & 1.522 & 0.089 \\ 
4 & $f_{0}(1710)$ & 8.30 & 1.676 & 0.163 \\ 
5 & $f_{2}(1270)$ & 3.16 & 1.290 & 0.196 \\ 
6 & $f_{2}(2150)$ & 2.22 & 2.162 & 0.159 \\ \hline
 & \textbf{SF} & 115.0 & & \\ \hline
\end{tabular}
\label{tab-1}
\end{table}

\begin{figure}[htbp]
  \centering
  \includegraphics[width=0.6\textwidth]{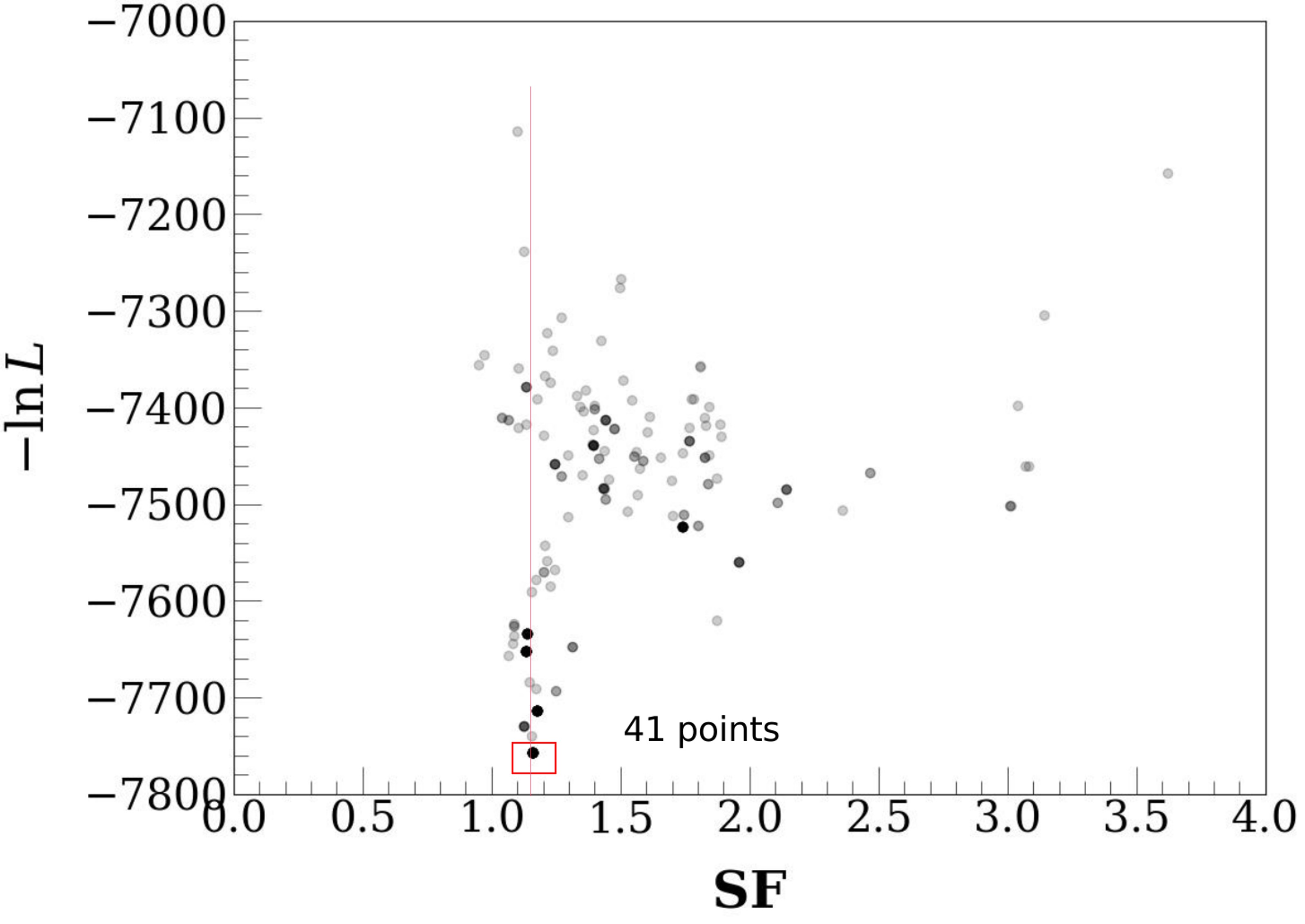}
  \caption{The distribution of likelihood values and $\mathbf{SF}$ values of the fitting results corresponding to the resonance combination $R_0$. The red vertical line represents the true value of $\textbf{SF}$, and the red box contains the points of good fits.}
  \label{fig-2}
\end{figure}

Firstly, we fit the MC sample $S_0$ with the $R_0$ model 300 times by using the target function $-\ln L$.  Figure~\ref{fig-2} shows the obtained logarithm of the likelihood and the sum of the fractions. It is apparently that even the fitting PWA model is perfectly matched to the data-producing model, there is still a large probability that the fitting results deviate significantly from the true values, while good fit results, in which the global minimum is found, always provide correct $\mathbf{SF}$ values. The red box of Fig.~\ref{fig-2} represents a region enclosing good fits. The number of points in it is $41$, that accounts for only about $14\%$ of the total fitting times. The unreliability of the fitting results is the so called non-convex problem, that is caused by the complexity of the PWA, resulting in various local minima of the likelihood function in the parameter space. One way to avoid this problem and find the global minima is by re-fitting data in huge number of times, with varied initial parameters, and this is a critical reason for the low efficiency of the PWA.

\begin{figure}[htbp]
  \centering
  \includegraphics[width=0.6\textwidth]{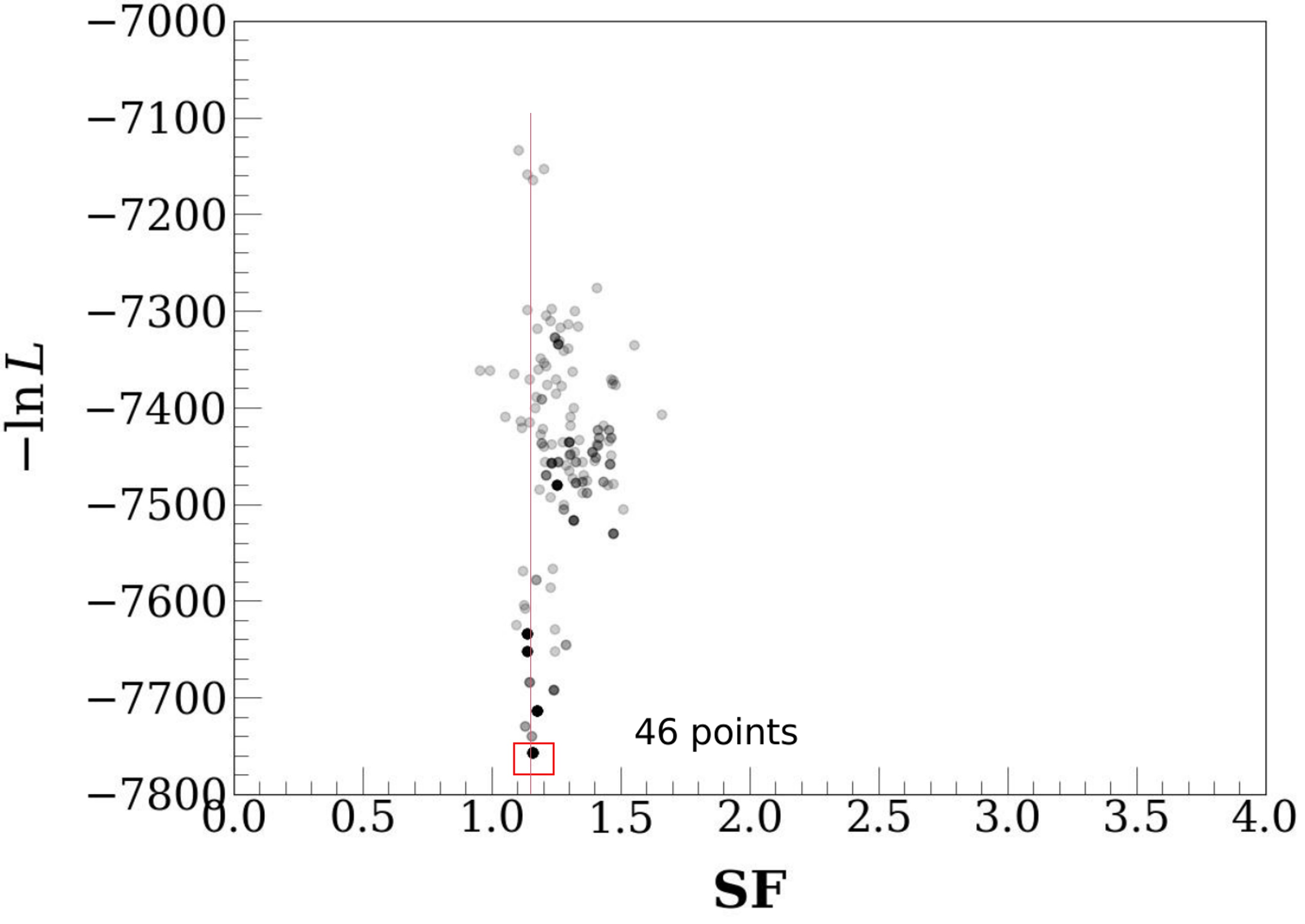} 
  \caption{The likelihood value and SF value distribution of the resonance state combination $R_0$ corresponding to the fitting result when $\mathbf{SF}=120\%$ and $\lambda=10^{-2}$. The red vertical line represents the true value of $\textbf{SF}$, and the red box contains the points of good fits.}
  \label{fig-3}
\end{figure}

Secondly, we redo the fits again by replacing the target function from $-\ln L$ to $\tilde{M}$. Usually, the expected sum of fractions $\overline{\mathbf{SF}}$ can be determined by a scanning method that will be described in Sec.~\ref{sec:scan_rss} along with the resonance selection. Here, we just adopt the result and set it to $120\%$, and set the strict-factor $\lambda = 10^{-2}$ by practical experience. The results of 300 fits are shown in Fig.~\ref{fig-3}. There are $46$ points in the red box of Fig.~\ref{fig-3}, which is slightly higher than the number in Fig.~\ref{fig-2}. It can be seen that the penalty term limits the range of $\mathbf{SF}$ as expected and increases the probability of the fitting result reaching the global optimum.

\begin{figure}[htbp]
  \centering
  \includegraphics[width=0.6\textwidth]{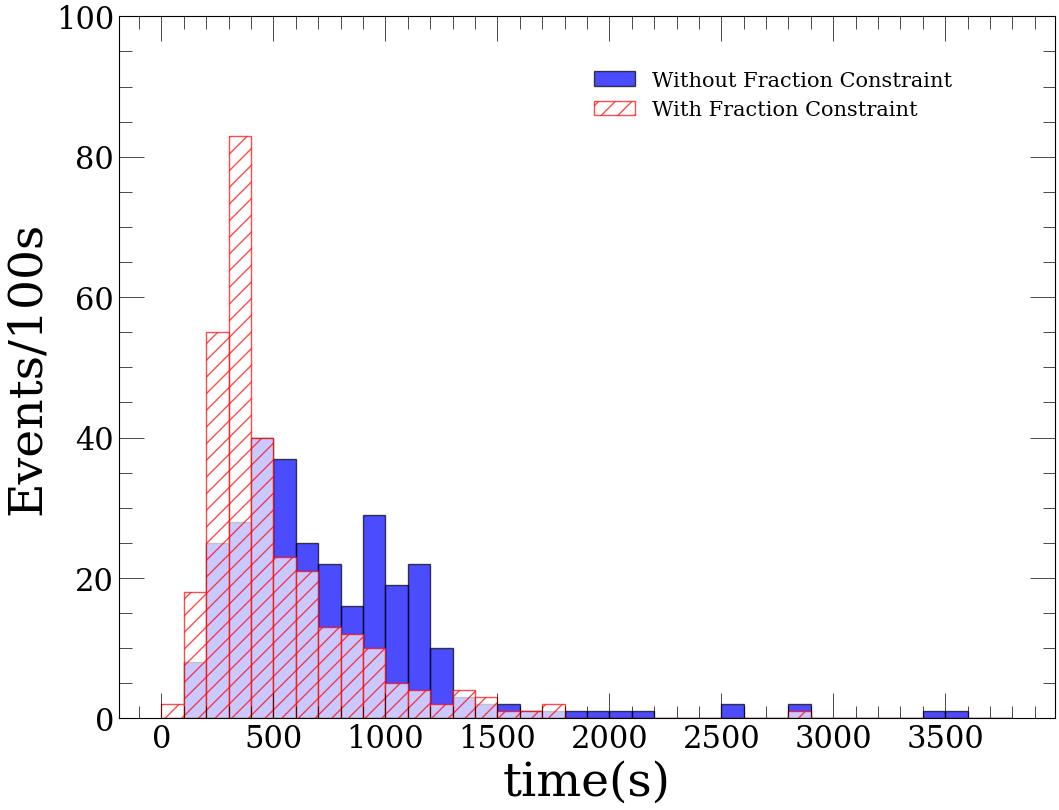} 
  \caption{Compare the fitting time used by $-\ln L$ and $\tilde{M}$.}
  \label{fig-4}
\end{figure}

Although it needs more computation source to calculate the penalty term $\mathbf{SF}$, against one's intuition, the whole fitting time required by $\tilde{M}$ is less than that of $-\ln L$. This timing reduction is mainly caused by the less tempts to find a minimal in a reduced parameter space. To investigate the impact on computation time, a time analysis is performed to obtain the results in Fig.~\ref{fig-2} and Fig.~\ref{fig-3}. The costumed time is shown in Fig.~\ref{fig-4}. From it, the average fitting time for $\tilde{M}$ is approximately $500$ s, while the average fitting time for $-\ln L$ is around $750$ s. A significant speed-up is found.  This result is obtained in our own testing environment, and factors such as the PWA program, fitting method, and hardware platform can affect the results. However, just like the role of penalty terms in the field of deep learning, the inclusion of penalty terms in this context serves to prevent large, ineffective attempts during the fitting process. These penalty terms provide additional gradients (on the boundaries of the parameter space) that are independent of the program, software, and hardware platforms used.

To check the feasibility of the new PWA method, the fitting results corresponding to the global optimal points, without or with the penalty, are listed in Table~\ref{tab-2} and Table~\ref{tab-3} for comparison. It can be seen that the two fitting results, including both mean values and statistical uncertainties, are consistent with each other. 

\begin{table}[htbp]
\caption{Fitting results of the PWA model $R_0$ with $-\ln L$.}
\centering
\begin{tabular}{|c|c|c|c|c|}
\hline
$R_0$ & Name & $F_i$(\%) & Mass (GeV) & Width (GeV) \\ \hline
1 & $f_{0}(980)$ & $39.2\pm1.5$ & $1.015\pm0.043$ & $0.102\pm0.030$ \\ 
2 & $f_{2}(2340)$ & $37.5\pm1.6$ & $2.571\pm0.015$ & $0.281\pm0.017$ \\ 
3 & $f^{'}_{2}(1525)$ & $23.5\pm1.0$ & $1.523\pm0.002$ & $0.084\pm0.003$ \\ 
4 & $f_{0}(1710)$ & $8.7\pm0.9$ & $1.671\pm0.005$ & $0.159\pm0.010$ \\ 
5 & $f_{2}(1270)$ & $2.7\pm0.6$ & $1.288\pm0.013$ & $0.181\pm0.027$ \\ 
6 & $f_{2}(2150)$ & $2.5\pm0.6$ & $2.152\pm0.012$ & $0.170\pm0.026$ \\ \hline
 & \textbf{SF} & $114.0$ & & \\ \hline
\end{tabular}
\label{tab-2}
\end{table}

\begin{table}[htbp]
\caption{Fitting results of the PWA model $R_0$ with $\tilde{M}$.}
\centering
\begin{tabular}{|c|c|c|c|c|}
\hline
$R_0$ & Name & $F_i$(\%) & Mass (GeV) & Width (GeV) \\ \hline
1 & $f_{0}(980)$ & $39.3\pm1.6$ & $1.017\pm0.039$ & $0.101\pm0.035$ \\ 
2 & $f_{2}(2340)$ & $37.5\pm1.8$ & $2.571\pm0.016$ & $0.282\pm0.018$ \\ 
3 & $f^{'}_{2}(1525)$ & $23.6\pm1.0$ & $1.523\pm0.002$ & $0.084\pm0.003$ \\ 
4 & $f_{0}(1710)$ & $8.7\pm 1.0$ & $1.671\pm0.005$ & $0.159\pm0.010$ \\ 
5 & $f_{2}(1270)$ & $2.7\pm0.6$ & $1.288\pm0.014$ & $0.182\pm0.026$ \\ 
6 & $f_{2}(2150)$ & $2.5\pm0.6$ & $2.152\pm0.012$ & $0.170\pm0.027$ \\ \hline
 & \textbf{SF} & $114.3$ & & \\ \hline
\end{tabular}
\label{tab-3}
\end{table}

\begin{figure}[htbp]
  \centering
  \includegraphics[width=0.6\textwidth]{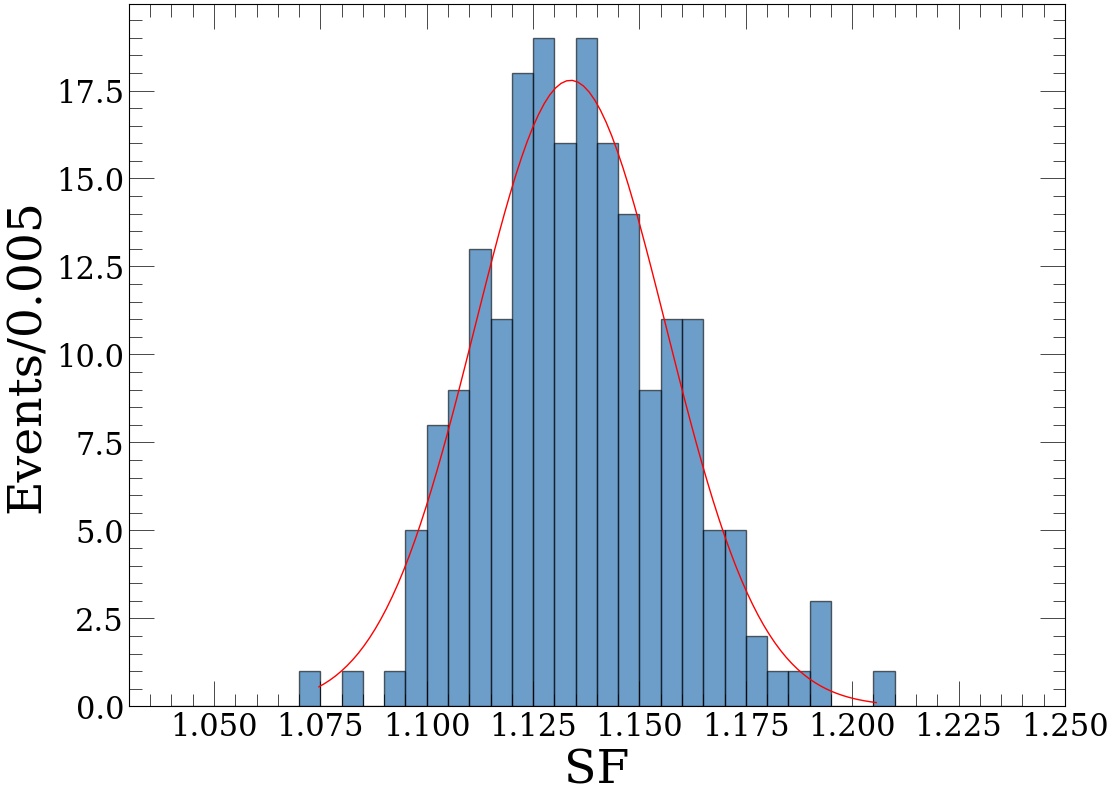} 
  \caption{The distribution of the sum of fractions in 300 test MC samples that are generated with the model $R_0$. The red curve represents the Gaussian function utilized in the fit.}
  \label{fig:test2}
\end{figure}

To test the fit stability of the PWA with the additional penalty term, we have generated 300 sets of samples using the same $R_0$ model only with various random number seeds, and performed fitting on each set. Figure~\ref{fig:test2} shows the distribution of the sum of fractions. A fit with a Gaussian function gives the result is $1.13\pm 0.02$, that is consistent with the input value $1.14$ considering the uncertainty.

\section{Fraction Constraint Scanning and Resonant State Selection}
\label{sec:scan_rss}
In the last section, both PWAs are performed with a perfect model, that is, exactly the one used in generating the MC sample. However, in a real PWA, to determine which resonances should be included is an important and difficult issue to be addressed~\cite{guegan2015model}. Typically, this is done by comparing the likelihood values of different combinations of resonances and calculating corresponding significance. But how to determine a baseline, that is crucial for the significance calculation, is a frequently debated question in PWA. Furthermore, whether to include a resonance or not should be beyond the sole goodness of a fit. In addition to considering the significance of a resonance, more information, such as the branching fraction, physical rules conservation, complexity of a PWA model, etc., need to be considered. Some researchers have already borrowed some mature theories from information theory, such as AIC and BIC~\cite{lasso}, to balance the model complexity and goodness of a fit.

\begin{figure}[htbp]
  \centering
  \includegraphics[width=0.6\textwidth]{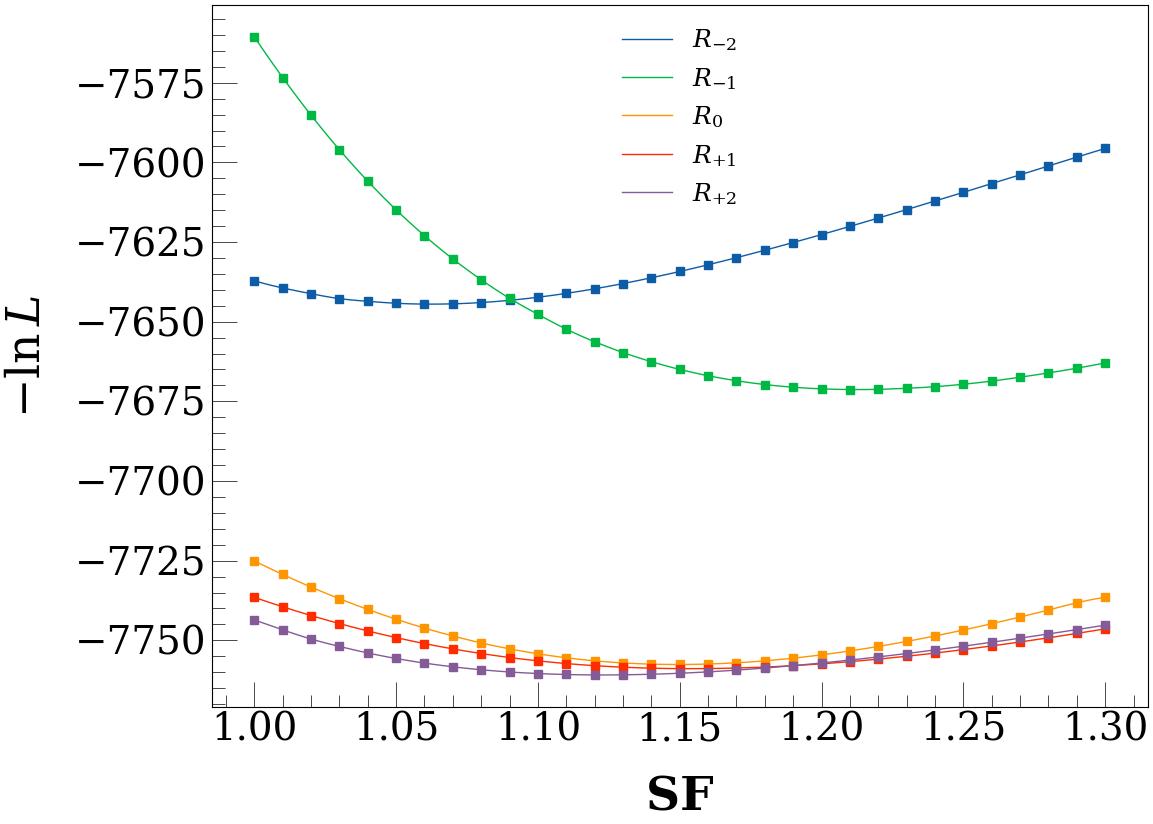} 
  \caption{$\mathbf{SF}$ scanning curves. The blue, green, yellow, red, and purple lines represent the PWA models $R_{-2}$, $R_{-1}$, $R_{0}$, $R_{+1}$, $R_{+2}$, respectively.}
  \label{fig-6}
\end{figure}

Similar to AIC and BIC, the fraction constraint method, proposed by us, try to control the model complexity by introducing the penalty term. Using $\tilde{M}$, we can quickly obtain the best fit results for different PWA models with various resonance combinations, when the strict-factor $\lambda$ is set to be a somewhat large value, such as $10^2$. Based on this advantage, the value of $\overline{\mathbf{SF}}$ is obtained by scanning in a series of fits, and the results are shown in Fig.~\ref{fig-6}. Here $R_{-1}$ represents the PWA model subtracting resonance $f_2(1270)$ from $R_0$, and $R_{-2}$ represents subtracting resonance $f_2(2150)$; while $R_{+1}$ represents adding resonance $f_0(1370)~$\cite{Bugg:1996ki}, $R_{+2}$ represents adding resonance  $f_2(2010)$\cite{Vladimirsky:2006ky}. 

From Fig.~\ref{fig-6}, it can be seen that there is a large gap between $R_{-1}$ ($(R_{-2}$) and $R_0$. The difference in the y-axis, i.e., the logarithm of the likelihood, indicates the models with subtracting resonances is not complex enough to describe the data, compared with the $R_0$. But the gap between $R_{+1}$ ($(R_{+2}$) and $R_0$ is very small, indicating that the parameters of models with additional resonances are overpopulated. Therefore, $R_0$ is the best PWA model to describe the data. So the scan method can help to select a reasonable set of resonances in a PWA model. And from the scan curve the best $\mathbf{SF}$ can be determined from the minimum, that should be considered as the expected value of $\mathbf{SF}$. 

\section{Summary}
This article proposes the use of $\tilde{M}$ instead of $-\ln L$ in PWA by evaluating the likelihood value as a function of fraction constraints, thereby improving analysis efficiency. An analysis conducted on the MC sample demonstrates the reliability of the fitted center values and statistical uncertainties based on the new method. Additionally, the relationship between the likelihood value of the fitting results and the $\mathbf{SF}$ value provides a fresh perspective on addressing the resonance selection issue. By constraining the $\mathbf{SF}$ values, redundant resonances can be effectively reduced, thereby mitigating the overestimation of systematic uncertainties resulting from the selection of resonance states. While the use of $\tilde{M}$ instead of $-\ln L$ does not offer a definitive solution to the increasingly complex nature of PWA driven by expanding data volumes, it has proven to enhance efficiency and minimize debates surrounding resonance states through practical implementation.

\bibliographystyle{unsrt}

\bibliography{sample}

\end{document}